\documentstyle[12pt]{article}
\newcommand{\NP}{Nucl. Phys. }
\newcommand{\NJP}{New J. Phys.}
\newcommand{\PR}{Phys. Rev. }
\newcommand{\PRL}{Phys. Rev. Lett. }
\newcommand{\PL}{Phys. Lett. }
\newcommand{\RMP}{Rev. Mod. Phys. }

\begin{document}
\baselineskip=16pt

\pagenumbering{arabic}

\vspace{1.0cm}
\begin{flushright}
LU-ITP 2005/015\\
hep-ph/0504018
\end{flushright}


\begin{center}
{\Large\sf Lepton mixing matrix in standard model extended by one sterile
neutrino}
\\[10pt]
\vspace{.5 cm}

{Yi Liao}
\vspace{1.0ex}

{\small Institut f\"ur Theoretische Physik, Universit\"at Leipzig,\\
Augustusplatz 10/11, D-04109 Leipzig, Germany\\}

\vspace{2.0ex}

{\bf Abstract}

\end{center}
We consider the simplest extension of the standard electroweak model by one
sterile neutrino that allows for neutrino masses and mixing. We find that its
leptonic sector contains much less free physical parameters than previously
realized. In addition to the two neutrino masses, the lepton mixing matrix
in charged current interactions involves $(n-1)$ free physical mixing angles
for $n$ generations. The mixing matrix in neutral current interactions of
neutrinos is completely fixed by the two masses. Both interactions conserve CP.
We illustrate the phenomenological implications of the model by vacuum neutrino
oscillations, tritium $\beta$ decay and neutrinoless double $\beta$ decay.
It turns out that, due to the revealed specific structure in its mixing matrix,
the model with any $n$ generations cannot accommodate simultaneously the data by
KamLAND, K2K and CHOOZ.

\begin{flushleft}


\end{flushleft}

\vspace{2cm}
\begin{center}
{\sf liaoy@itp.uni-leipzig.de}
\end{center}

\newpage

\section{Introduction}
The experiments with solar, atmospheric, reactor and accelerator neutrinos have
now provided evidence that neutrinos have mass and mix $\cite{review,kayser89}$.
These experimental results can be best understood in terms of neutrino
oscillations $\cite{nuoscill}$.
This implies unambiguously that the standard model (SM) of electroweak
interactions has to be extended at least in the leptonic sector.
Most phenomenological analyses performed so far assume three active
massive neutrinos and a three by three unitary mixing matrix for
charged current (CC) interactions of leptons.
The leptonic sector then contains two differences of neutrino masses
squared, three mixing angles plus three CP-violating phases. It has been
shown to be capable of accommodating all neutrino data except the LSND
result $\cite{lsnd}$, which however was not confirmed by other short baseline
experiments $\cite{others}$ and remains to be clarified by
MiniBooNE $\cite{boone}$ in the future.
From the model building point of view, such a scenario requires new degrees of
freedom to be added to SM, either some heavy neutrinos $\cite{seesaw}$,
or a Higgs triplet which also develops a vacuum expectation value
$\cite{triplet}$. In the first case, the three by three
mixing matrix amongst three light, active neutrinos is only approximately unitary
to the extent that their mixing with heavy neutrinos can be ignored in the
analysis of current data. Thus it amounts to a low energy effective theory of a
fundamental one which could be much more complicated. In the second one,
the existence of a non-doublet Higgs boson is always severely constrained by
precision electroweak data and the null result of direct Higgs searches.

There are also attempts to incorporating the LSND result by including explicitly
a sterile neutrino into the mixing scheme which was introduced earlier in the other
context $\cite{sterile}$.
But they are found to be disfavored by the experimental data either because of
the tension between the positive result of LSND and the negative ones by other
short baseline experiments or because of the rejection of sterile neutrino's
involvement in solar and atmospheric data $\cite{4nu}$.

In this work, we put aside the LSND result as in most studies and ask whether
it is possible to understand the neutrino data in a minimal extension of SM.
The extension is minimal in the sense that it introduces the least numbers of
new degrees of freedom and free physical parameters into a fundamental theory
without endangering precision electroweak data. With this in mind, one
possibility would be to introduce one sterile neutrino to the SM of three
generations. Such kind of models were systematically studied long time ago in
the pioneering work of Ref.$\cite{sv80a}$. They were also considered without
including Majorana mass terms in Ref.$\cite{sv80b}$.
According to the analysis in Ref.$\cite{sv80a}$, this would introduce 5
mixing angles and 3 CP-violating phases, in addition to two neutrino masses
(with the other two being massless). There should thus be some room to
accommodate the mentioned neutrino data that essentially call for two
mass-squared differences and 3 independent mixing angles.
However, as we shall analyse in the next section, there are actually only 2
mixing angles and no CP-violating phase in the lepton mixing matrix after
we take into account completely the constraints from the texture zero that
appears in the original neutrino mass matrix $\cite{pilaftsis}$.
Furthermore, as we shall show in section 3, due to the special structure
in the mixing matrix exposed in section 2, it is even not possible to
accommodate simultaneously the data by KamLAND, K2K and CHOOZ
$\cite{kamland,k2k,chooz}$ for any number
of $n$ generations, although in that case there are $(n-1)$ free mixing angles
at our disposal. The same structure also results in the vanishing leading
contribution to neutrinoless double $\beta$ decay for any $n$ which in
principle is allowed to occur due to the Majorana nature of the neutrinos.
Our results are summarized and conclusions are made in the last section.

\section{Parametrization of mixing matrix}
In this section we first describe the leptonic sector of the $n$-generation
SM extended by $n_0$ sterile neutrinos. Then, we specialize to the simplest
case of $n_0=1$ to parametrize the lepton mixing matrix and count its
independent physical parameters.

\subsection{The model}
The only new fields compared to SM are the $n_0$ sterile neutrinos that we
choose to be right-handed without loss of generality,
$s_{Rx}$, $x=1,\cdots,n_0$.
It is sufficient for us to concentrate on the leptonic sector of the model
that contains the standard $n$ generations of the doublets,
$F_{La}=(n_{La},f_{La})^T$,
and of the charged lepton singlets,
$f_{Ra}$, $a=1,\cdots,n$.
Here $L,R$ refer to the left- and right-handed projections of the fields in
terms of $P_{L,R}=(1\mp\gamma^5)/2$. The kinetic and gauge interaction terms
of the leptons are
\begin{equation}
\begin{array}{rcl}
{\cal L}_{\rm k}&=&\displaystyle
\overline{n_{La}}i\rlap/\partial n_{La}
+\overline{s_{Rx}}i\rlap/\partial s_{Rx}
+\overline{f_a}i\rlap/\partial f_a\\
{\cal L}_{\rm CC}&=&\displaystyle
\frac{g}{\sqrt{2}}\left[W^+_{\mu}\overline{n_{La}}\gamma^{\mu}f_{La}
+W^-_{\mu}\overline{f_{La}}\gamma^{\mu}n_{La}\right]\\
{\cal L}_{\rm NC}&=&\displaystyle
\frac{g}{2c_W}Z_{\mu}\left[\overline{n_{La}}\gamma^{\mu}n_{La}
+\overline{f_a}\gamma^{\mu}\left(-P_L+2s_W^2\right)f_a\right]\\
{\cal L}_{\rm EM}&=&\displaystyle
-eA_{\mu}\overline{f_a}\gamma^{\mu}f_a
\end{array}
\end{equation}
where summation over indices $a,x$ is implied. $g,~e$ are respectively
the couplings of $SU(2)_L$ and $U(1)_{\rm EM}$, and $c_W=m_W/m_Z$ with
$m_{W,Z}$ being the masses of $W^{\pm},~Z$.

Since the sterile neutrinos are neutral under $SU(2)_L\times U(1)_Y$ by
definition, they can have bare mass terms of Majorana type,
\begin{equation}
-{\cal L}_{s_R}=\frac{1}{2}M_{xy}\overline{s_{Rx}^C}s_{Ry}
+\frac{1}{2}M^*_{xy}\overline{s_{Ry}}s_{Rx}^C
\end{equation}
where $\psi^C={\cal C}\gamma^0\psi^*$ stands for the charge-conjugate field of
$\psi$ with ${\cal C}=i\gamma^0\gamma^2$ satisfying
${\cal C}=-{\cal C}^{\dagger}=-{\cal C}^T=-{\cal C}^{-1}$ and
${\cal C}\gamma^{\mu T}{\cal C}=\gamma^{\mu}$. We denote $s_R^C=(s_R)^C$ for
brevity. The $n_0\times n_0$ complex matrix $M$ is symmetric due to
anticommutativity of fermion fields, but is otherwise general.
Together with the Yukawa terms of the leptons,
\begin{equation}
-{\cal L}_{\rm Y}=y_{ab}^f\overline{F_{La}}\varphi f_{Rb}
+y_{ax}^n\overline{F_{La}}\tilde{\varphi}s_{Rx}+{\rm h.c.}
\end{equation}
where $\varphi$ is the Higgs doublet field that develops a vacuum expectation
value, $\langle\varphi\rangle=(0,1)^Tv/\sqrt{2}$, and
$\tilde{\varphi}=i\sigma^2\varphi^*$, the lepton mass terms become
\begin{equation}
-{\cal L}_{\rm m}=
\left[\overline{f_L}m^f f_R+\overline{n_L}Ds_R
+{\rm h.c.}\right]
+\frac{1}{2}\left[\overline{s_R^C}Ms_R+{\rm h.c.}\right]
\end{equation}
where $m^f=y^fv/\sqrt{2}$ and $D=y^nv/\sqrt{2}$ are $n\times n$ and
$n\times n_0$ complex matrices respectively. The charged lepton mass terms are
diagonalized as usual by biunitary transformations,
\begin{equation}
f_L=X_L\ell_L,~f_R=X_R\ell_R,~X_{L,R}^{-1}=X_{L,R}^{\dagger},
\end{equation}
with
\begin{equation}
X_L^{\dagger}m^fX_R=m^{\ell}={\rm diag}(m_e,m_{\mu},m_{\tau},\cdots)
\end{equation}
being real and positive. We shall denote the mass eigenstate fields of the
charged leptons, $\ell_{L,R}$, by the Greek indices,
$\alpha,\beta=1,2,\cdots,n$, corresponding to the electron, muon, etc.

The kinetic and mass terms for the neutral leptons
contain the fields $n_L,~s_R$ and $s_R^C$. To diagonalize, we first rewrite
them uniformly in terms of the fields $n_L,~s_R$ and their charge-conjugates
$n_L^C,~s_R^C$. Using
$\overline{\psi^C}i\gamma^{\mu}\partial_{\mu}\chi^C
=-i\partial_{\mu}(\bar{\chi}\gamma^{\mu}\psi)
+\bar{\chi}i\gamma^{\mu}\partial_{\mu}\psi$,
where the total derivative term can be ignored in the Lagrangian, and
$\overline{\psi^C}\chi^C=\bar{\chi}\psi$, the terms become
\begin{equation}
\begin{array}{rcl}
{\cal L}_{\rm k}^{\nu}&=&\displaystyle
\frac{1}{2}\left(\overline{n_L^C},\overline{s_R}\right)
i\rlap/\partial\left(\begin{array}{c}n_L^C\\s_R
\end{array}\right)
+\frac{1}{2}\left(\overline{n_L},\overline{s_R^C}\right)
i\rlap/\partial\left(\begin{array}{c}n_L\\s_R^C
\end{array}\right)\\
-{\cal L}_{\rm m}^{\nu}&=&\displaystyle
\frac{1}{2}\left(\overline{n_L},\overline{s_R^C}\right)m^n
\left(\begin{array}{c}n_L^C\\s_R
\end{array}\right)+
\frac{1}{2}\left(\overline{n_L^C},\overline{s_R}\right)m^{n\dagger}
\left(\begin{array}{c}n_L\\s_R^C
\end{array}\right)
\end{array}
\end{equation}
where the $(n+n_0)$ dimensional, symmetric mass matrix in the new basis is
\begin{equation}
m^n=\left(\begin{array}{cc}0_n&D\\D^T&M
\end{array}\right)
\label{eq_mass}
\end{equation}
with $0_n$ being the zero matrix of $n$ dimensions. For $n>n_0$, which covers
our interested case $n=3,~n_0=1$ later on, it contains a zero eigenvalue of
degeneracy $(n-n_0)$ and $2n_0$ eigenvalues which are non-zero and nondegenerate
for general parameters $D,~M$. Without changing the diagonal form of the
kinetic terms, we make a unitary transformation
\begin{equation}
\left(\begin{array}{c}n_L^C\\s_R
\end{array}\right)=Y\nu_R,~Y^{-1}=Y^{\dagger},
\end{equation}
which also fixes the transformation of the conjugate fields,
\begin{equation}
\left(\begin{array}{c}n_L\\s_R^C
\end{array}\right)=Y^*\nu_R^C,
\end{equation}
such that
\begin{equation}
Y^Tm^nY=m^{\nu}={\rm diag}(0,\cdots,0,m_{n-n_0+1},\cdots,m_{n+n_0})
\label{eq_diag}
\end{equation}
with the nonvanishing masses being real and positive. We shall denote the
mass eigenstate fields of the neutral leptons, $\nu_R$, by the Latin indices,
$j,k=1,2,\cdots,n+n_0$. Then,
\begin{equation}
{\cal L}^{\nu}_{\rm k}+{\cal L}^{\nu}_{\rm m}=
\frac{1}{2}\left(\overline{\nu_R}i\rlap/\partial \nu_R
+\overline{\nu_R^C}i\rlap/\partial \nu_R^C\right)
-\frac{1}{2}\left(\overline{\nu_R^C}m^{\nu}\nu_R
+\overline{\nu_R}m^{\nu}\nu_R^C\right)
\end{equation}
which may be put in the compact form
\begin{equation}
{\cal L}^{\nu}_{\rm k}+{\cal L}^{\nu}_{\rm m}
=\frac{1}{2}\bar{\nu}\left(i\rlap/\partial-m^{\nu}\right)\nu
\end{equation}
by introducing the Majorana neutrino fields
\begin{equation}
\nu=\nu_R+\nu_R^C
\end{equation}
satisfying $\nu^C=\nu$.

Now we express the interactions of leptons in terms of the fields with a
definite mass. There are no changes in
${\cal L}_{\rm NC}^{\ell}+{\cal L}_{\rm EM}^{\ell}$
for the charged leptons. The CC interaction becomes
\begin{equation}
{\cal L}_{\rm CC}=\displaystyle\frac{g}{\sqrt{2}}\left[
V^{\rm C}_{\beta j}W^-_{\mu}\overline{\ell_{L\beta}}\gamma^{\mu}\nu_j
+V^{{\rm C}*}_{\beta j}W^+_{\mu}\overline{\nu_j}\gamma^{\mu}\ell_{L\beta}
\right]
\end{equation}
where the $n\times (n+n_0)$ matrix
\begin{equation}
V^{\rm C}_{\beta j}=\sum_{a=1}^n(X^{\dagger}_L)_{\beta a}Y^*_{a j},
\label{eq_vc}
\end{equation}
is the leptonic analog of the mixing matrix $V^{\dagger}_{\rm CKM}$ in the
hadronic sector. It is important to notice that only the first $n$ rows in $Y$
are involved in the CC mixing matrix $V^{\rm C}$ $\cite{sv80a}$ because the
remaining $n_0$ rows are associated with the sterile neutrinos $s_R$ which do
not enter any interactions. Due to this and unitarity of $X_L$ and $Y$,
we have
\begin{equation}
V^{\rm C}V^{{\rm C}\dagger}=1_n,
\label{eq_unit}
\end{equation}
but $V^{{\rm C}\dagger}V^{\rm C}\ne 1_{n+n_0}$. Actually, the latter appears
in the neutral current (NC) interaction for the neutrinos
\begin{equation}
{\cal L}_{\rm NC}^{\nu}=\displaystyle
\frac{g}{2c_W}V^{\rm N}_{kj}Z_{\mu}\overline{\nu_{k}}\gamma^{\mu}P_L\nu_{j}
\end{equation}
by the relation $\cite{sv80a}$
\begin{equation}
V^{\rm N}_{kj}=\sum_{a,b=1}^nY_{bk}\delta_{ba}Y^*_{aj}
=\sum_{a,b=1}^n\sum_{\alpha}
Y_{bk}(X_L)_{b\alpha}(X_L^{\dagger})_{\alpha a}Y^*_{aj}
=(V^{{\rm C}\dagger}V^{\rm C})_{kj}
\end{equation}
Using
$\overline{\psi^C}\gamma^{\mu}P_L\chi^C=-\bar{\chi}\gamma^{\mu}P_R\psi$,
$\nu_j^C=\nu_j$, and Hermiticity of $V^{\rm N}$, the interaction can also be
cast in the form,
\begin{equation}
{\cal L}_{\rm NC}^{\nu}=\frac{g}{4c_W}Z_{\mu}
\overline{\nu_k}\gamma^{\mu}\left[i{\rm ~Im~}V^{\rm N}_{kj}
-\gamma^5{\rm ~Re~}V^{\rm N}_{kj}\right]\nu_j
\end{equation}

In addition to the condition $(\ref{eq_unit})$, $V^{\rm C}$
satisfies a relation that will be important in its parametrization. The original
neutrino mass matrix $m^n$ has an $n$ dimensional zero submatrix in its
left-upper corner which is protected by gauge symmetry of the model. Then,
eq. $(\ref{eq_diag})$ implies that
\begin{equation}
(Y^*m^{\nu}Y^{\dagger})_{ab}=0,~a,b=1,\cdots,n
\end{equation}
Multiplying it by $(X_L^{\dagger})_{\alpha a}(X_L^{\dagger})_{\beta b}$, summing
over $a,~b$ and using eq. $(\ref{eq_vc})$ leads to the matrix relation
$\cite{pilaftsis}$
\begin{equation}
V^{\rm C}m^{\nu}V^{{\rm C}T}=0_n
\label{eq_zero}
\end{equation}
Note that the above holds irrespective of $n>n_0$ when there are massless modes
or $n\le n_0$ when there is none for general parameters $D,~M$.

\subsection{The mixing matrix}

From now on, we shall restrict ourselves to the case of $n_0=1$. We
offer two ways to construct the mixing matrix $V^{\rm C}$, one by using the
constraints $(\ref{eq_unit},\ref{eq_zero})$ and the other by explicit
diagonalization. This is then followed by counting the independent physical
parameters contained in it.

We start with the case of $n=1$ in which the two neutrinos are generally massive
and nondegenerate with masses denoted by $0<m_-<m_+$. The mixing matrix is a
one-row complex matrix, $V^{\rm C}=(r_1e^{i\rho_1},~r_2e^{i\rho_2})$.
Eqs. $(\ref{eq_unit},\ref{eq_zero})$ read
\begin{equation}
r_1^2+r_2^2=1,~m_-r_1^2+m_+r_2^2e^{2i(\rho_2-\rho_1)}=0
\end{equation}
whose solution gives
\begin{equation}
V^{\rm C}=e^{i\rho_1}(c_m,\pm is_m),
~c_m=\sqrt{\frac{m_+}{m_++m_-}},~s_m=\sqrt{\frac{m_-}{m_++m_-}}
\label{eq_solu_1}
\end{equation}
and then,
\begin{equation}
V^{\rm N}=\left(\begin{array}{ll}c_m^2&\pm ic_ms_m\\\mp ic_ms_m&s_m^2
\end{array}\right)
\end{equation}
The global phase in $V^{\rm C}$ can be absorbed into the electron field while
the two signs can be interchanged by flipping the sign of the field $\nu_2$.
Both mixing matrices are thus uniquely fixed by the two neutrino
masses, in contrast to the claim in Ref.$\cite{sv80a}$ that there are one
free mixing angle and one free CP-violating phase.
Note that the nonreality of $V^{\rm C,N}$ is not a sign of CP violation.
If we want, we can absorb the $i$ into the field $\nu_2$ without changing
its mass term so that $V^{\rm C,N}$ are real. But this will
introduce a nontrivial ``creation phase'' into it which was unity before
rephasing $\cite{kayser}$.

The above result may also be obtained by direct diagonalization of the mass
matrix in eq. $(\ref{eq_mass})$ where $D$ and $M$ are two complex numbers.
The unitary matrix $Y$ in eq. $(\ref{eq_diag})$ may be parametrized generally
as
\begin{equation}
Y=e^{i\gamma_0}\left(\begin{array}{ll}
ce^{i\gamma_1}&se^{i\gamma_2}\\
-se^{-i\gamma_2}&ce^{-i\gamma_1}
\end{array}\right)
\label{eq_y}
\end{equation}
with $c^2+s^2=1$ and all parameters being real. Denoting
$D=|D|e^{i\delta_1}$, $M=|M|e^{i\delta_2}$, eq. $(\ref{eq_diag})$ then yields
\begin{equation}
\begin{array}{l}
\displaystyle
m_{\pm}=\frac{1}{2}\left[\sqrt{|M|^2+4|D|^2}\pm|M|\right]\\
e^{i2\gamma_1}=-e^{i2\gamma_2}=\pm ie^{-i(\delta_1-\delta_2)},
~e^{i2\gamma_0}=\pm ie^{-i\delta_1}
\end{array}
\label{eq_gamma}
\end{equation}
and $c,~s$ being identified with $c_m,~s_m$ in eq. $(\ref{eq_solu_1})$.
Since only the first row of $Y$ appears in $V^{\rm C}$, the relevant phase is
$e^{i(\gamma_2-\gamma_1)}=\pm i$ independently of $\delta_{1,2}$, and the
preceding result is reproduced.

For the case of $n>n_0=1$, we generally have $2$ massive modes with masses
$m_-<m_+$ that we arrange
to be the last two in the order of increasing mass and $(n-1)$ massless modes.
Assuming $(V^{\rm C})_{\alpha,j}=r_{\alpha,j}e^{i\rho_{\alpha,j}}$,
eq. $(\ref{eq_zero})$ implies that
\begin{equation}
m_-r_{\alpha,n}r_{\beta,n}e^{i(\rho_{\alpha,n}+\rho_{\beta,n})}
+m_+r_{\alpha,n+1}r_{\beta,n+1}e^{i(\rho_{\alpha,n+1}+\rho_{\beta,n+1})}=0
\end{equation}
Note that the constraints from the off-diagonal elements ($\alpha\ne\beta$) are
not independent but just resolve the separate two-fold ambiguities in the
constraints for the diagonal elements ($\alpha=\beta$) to an overall two-fold
ambiguity:
\begin{equation}
\displaystyle\frac{r_{\alpha,n+1}}{r_{\alpha,n}}=\sqrt{r_m},
~e^{i\rho_{\alpha,n+1}}=\pm ie^{i\rho_{\alpha,n}}
\end{equation}
with $\displaystyle r_m=\frac{m_-}{m_+}$. $V^{\rm C}$ can then be factorized as
\begin{equation}
V^{\rm C}=VU
\label{eq_factor}
\end{equation}
where $V$ is an $n\times n$ matrix whose $\alpha$-th row is
\begin{equation}
\left(r_{\alpha,1}e^{i\rho_{\alpha,1}},\cdots,r_{\alpha,n-1}e^{i\rho_{\alpha,n-1}},
r_{\alpha,n}\sqrt{1+r_m}e^{i\rho_{\alpha,n}}\right)
\end{equation}
and $U$ is an $n\times (n+1)$ matrix
\begin{equation}
U=\left(
\begin{array}{lll}1_{n-1}&0_{(n-1)\times 1}&0_{(n-1)\times 1}\\
0_{1\times(n-1)}&c_m&\pm is_m
\end{array}\right)
\label{eq_u}
\end{equation}
with $c_m,~s_m$ again given in eq. $(\ref{eq_solu_1})$.
$UU^{\dagger}=1_n$ together with eq. $(\ref{eq_unit})$ gives,
$VV^{\dagger}=1_n$. Since $V$ is a square matrix whose entries are otherwise
arbitrary, it is a general unitary matrix of dimension $n$. The above structure
yields
\begin{equation}
V^{\rm N}=\left(
\begin{array}{lll}1_{n-1}&&\\&c_m^2&\pm ic_ms_m\\&\mp ic_ms_m&s_m^2
\end{array}\right)
\end{equation}
Thus the NC interactions of neutrinos do not contain any free
mixing angles or CP-violating phases that may appear in their CC
interactions. They conserve CP, and the off-diagonal interactions occur only
between the two massive neutrinos.

This result can be confirmed by direct diagonalization of the mass matrix in
eq. $(\ref{eq_mass})$. In the first step, we decouple the $(n-1)$ massless
modes. Now $D$ is an $n$-row column matrix whose entries are
parametrized as $D_a=|D_a|e^{i\delta_a}$. The magnitudes define a vector
in ${\cal R}^n$, which is rotated to the $n$-th axis by the rotation $R_0$.
Denoting $E_0={\rm diag}(e^{-i\delta_1},\cdots,e^{-i\delta_n})$ and the $(n+1)$
dimensional unitary matrix,
\begin{equation}
Y_0=\left(\begin{array}{cc}(R_0E_0)^T&\\&1
\end{array}\right)
\end{equation}
we have
\begin{equation}
Y^T_0m^nY_0=\left(\begin{array}{ccc}0_{n-1}&&\\&0&|D|\\&|D|&M
\end{array}\right)
\end{equation}
with $|D|=\sqrt{\sum_a|D_a|^2}$. The problem has thus been reduced to the case
of $n=n_0=1$ that we treated earlier. The required two-dimensional unitary
matrix, now denoted as $y_1$, for diagonalizing the submatrix is given
in eqs. $(\ref{eq_y},\ref{eq_gamma})$. Then,
\begin{equation}
Y^Tm^nY={\rm diag}(0,\cdots,0,m_-,m_+)
\end{equation}
where $Y=Y_0Y_1$ and
\begin{equation}
Y_1=\left(\begin{array}{cc}1_{n-1}&\\&y_1
\end{array}\right)
\end{equation}
Noting the block diagonal form of $Y_{0,1}$, eq. $(\ref{eq_vc})$ gives the same
factorized form of $V^{\rm C}$ as in eq. $(\ref{eq_factor})$ with the same $U$
as in eq. $(\ref{eq_u})$ but now
\begin{equation}
V=X_L^{\dagger}(R_0E_0)^{\dagger}e^{i(\gamma_0+\gamma_1)}
\end{equation}
Since $X_L$ is a general unitary matrix, so is $V$.

Now we count the free physical parameters contained in $V^{\rm C}$. An $n$
dimensional unitary matrix $V$ may be parametrized as a product in any
arbitrarily specified order of the $n$ phase matrices,
$e_{\alpha}(u_{\alpha})$ ($\alpha=1,\cdots,n$),
and the $n(n-1)/2$ complex rotation matrices in the $(\alpha,\beta)$ plane,
$\omega_{\alpha\beta}(\theta_{\alpha\beta},\varphi_{\alpha\beta})$
($n\ge\beta>\alpha\ge 1$) $\cite{sv80a}$.
Here $e_{\alpha}(z)$ is obtained by replacing the $\alpha$-th entry
in $1_n$ by the phase $z$, and
\begin{equation}
\omega_{\alpha\beta}(\theta_{\alpha\beta},\varphi_{\alpha\beta})
=e_{\alpha}(e^{i\varphi_{\alpha\beta}})R_{\alpha\beta}(\theta_{\alpha\beta})
e_{\alpha}(e^{-i\varphi_{\alpha\beta}})
\end{equation}
where $R_{\alpha\beta}(\theta_{\alpha\beta})$ is the usual real rotation matrix
through angle $\theta_{\alpha\beta}$ in the $(\alpha,\beta)$ plane.
We choose the order of products in such a way that it fits our purpose here:
\begin{equation}
V=e_n(u_n)\prod_{\alpha=1}^{n-1}\omega_{\alpha n}
(\theta_{\alpha n},\varphi_{\alpha n})
\left(\begin{array}{cc}X&\\&1
\end{array}\right)\equiv V_0\left(\begin{array}{cc}X&\\&1
\end{array}\right),
\end{equation}
where $X$ is the general $(n-1)$ dimensional unitary matrix containing $(n-1)^2$
real parameters and the $\omega_{\alpha n}$ factors are ordered from left to
right in increasing $\alpha$. From the revealed structure in
eq. $(\ref{eq_factor})$, the matrix containing $X$ can be pushed through the
matrix $U$ to be absorbed into the massless neutrino fields without causing any
other changes in the Lagrangian. This leaves us with $(2n-1)$ real parameters
in $V_0$.

However, not all parameters in $V_0$ are physical. For brevity, we denote
$\omega_{\alpha n}(\theta_{\alpha n},\varphi_{\alpha n})=\omega_{\alpha}$,
$R_{\alpha n}(\theta_{\alpha n})=R_{\alpha}$,
$e_{\alpha}(e^{i\varphi_{\alpha n}})=e_{\alpha}$, and
$e_{\alpha}(e^{-i\varphi_{\alpha n}})=e_{\alpha}^*$.
For $n=1$, $V_0=e_1(u_1)$ can be absorbed into the single charged lepton field
leaving no free physical parameters in $V^{\rm C}$ as we found earlier. For
$n=2$, we have
\begin{equation}
V_0=e_2(u_2)e_1R_1(\theta_1)e_1^*
\end{equation}
where $e_2(u_2)e_1$ can be absorbed into the two charged lepton fields
while $e_1^*$ can be pushed through $U$ and absorbed into the massless
neutrino field, leaving behind one physical mixing angle in this case.
For $n\ge 3$, we combine the two adjacent $\omega$'s as follows,
\begin{equation}
\omega_{\alpha}\omega_{\alpha+1}
=e_{\alpha}R_{\alpha}e_{\alpha}^*e_{\alpha+1}R_{\alpha+1}e_{\alpha+1}^*
=e_{\alpha}e_{\alpha+1}R_{\alpha}R_{\alpha+1}e_{\alpha}^*e_{\alpha+1}^*
\end{equation}
because
$[e_{\alpha},e_{\beta}]=0$, and $[e_{\alpha},R_{\beta}]=0$ for
$\alpha\ne\beta$ and $\alpha<n$. The sequence can be continued such that
\begin{equation}
V_0=\left[e_n(u_n)\prod_{\alpha=1}^{n-1}e_{\alpha}\right]\cdot
\prod_{\beta=1}^{n-1}R_{\beta}\cdot
\left[\prod_{\gamma=1}^{n-1}e_{\gamma}^*\right]
\end{equation}
Again, the right factor commutes with $U$ to get absorbed by the $(n-1)$
massless neutrino fields and the left one by the $n$ charged lepton fields.

To summarize, the lepton mixing matrix $V^{\rm C}$ contains $(n-1)$ free
physical mixing angles and no CP-violating phases, and may be parametrized as
in eq. $(\ref{eq_factor})$ with $U$ given in eq. $(\ref{eq_u})$ and
\begin{equation}
V=\prod_{\alpha=1}^{n-1}R_{\alpha n}(\theta_{\alpha})
\label{eq_v}
\end{equation}
The angles $\theta_{\alpha}$ describe the mixing of the $(n-1)$ massless
neutrinos with the subsystem of the two massive ones. The mixing matrix $V^{\rm N}$
in NC interactions of neutrinos does not contain these free
parameters but is fixed by the two neutrino masses. It is off-diagonal
only in the massive subsystem and also conserves CP.
It is interesting that the above counting for $V^{\rm C}$ happens to be the same
as the one in Ref.$\cite{sv80b}$ where the bare Majorana mass terms were not
included. In that case, mixing occurs only amongst the $n$ left-handed neutrinos
that belong to the leptonic doublets, and there are one massive Dirac neutrino
and $(n-1)$ massless neutrinos. Since a massive Dirac neutrino may be considered
as a pair of Majorana neutrinos with identical mass, the model studied in
Ref.$\cite{sv80b}$ appears as a special case in the current work. The Majorana
mass terms were indeed included in Ref.$\cite{sv80a}$, but it was found that
there are $(2n-1)$ mixing angles and $n$ CP-violating phases in $V^{\rm C}$,
totaling $(3n-1)$ free parameters, much more than the number found here. The
difference arises from the fact that the constraints $(\ref{eq_zero})$ from
the texture zero in the original neutrino mass matrix have been completely
exploited here to remove all unphysical parameters while they were only partially
applied in Ref.$\cite{sv80b}$ to delete unitary transformations within the
massless neutrinos.

The above counting of physical parameters can be extended to the general case.
Without giving further details, we record the results as follows.
For  $n_0\ge n\ge 1$, $V^{\rm C}$ contains $n(n_0-1)$ mixing angles and
$n(n_0-1)$ CP phases, to be compared with $n_0n+n(n-1)/2$ angles and
$n_0n+n(n-1)/2$ phases in Ref. $\cite{sv80a}$. Out of them, only $n(2n_0-n-1)/2$
angles and $n(2n_0-n-1)/2$ phases appear in $V^{\rm N}$.
When $n\ge n_0\ge 1$, $V^{\rm C}$ has $n_0(n-1)$ angles and $n(n_0-1)$ phases,
much less than $2n_0n-n_0(n_0+1)/2$ angles and $2n_0n-n-n_0(n_0-1)/2$ phases
found in Ref. $\cite{sv80a}$. Out of those, only $n_0(n_0-1)/2$ angles and
$n_0(n_0-1)/2$ phases enter in $V_{\rm N}$.

\section{Phenomenological implications}

We study in this section some phenomenological implications of the leptonic
CC interactions as revealed in eqs. $(\ref{eq_factor},\ref{eq_u},\ref{eq_v})$.
A possible way to measure the absolute neutrino mass is to study the electron
spectrum in tritium $\beta$ decay, which is sensitive to the effective
neutrino mass
\begin{equation}
m_{\nu_e}=\sqrt{\sum_jm_j^2|V^{\rm C}_{1j}|^2}
=|V_{1n}|\sqrt{m_-^2c_m^2+m_+^2s_m^2}
=|V_{1n}|\sqrt{m_+m_-}=|V_{1n}||D|
\end{equation}
Thus the decay spectrum is only sensitive to Dirac-type mass in this model.

The neutrinoless double $\beta$ decay of nuclei is currently the only known
practical means to unravel the Majorana nature of neutrinos.
At the leading order of expansion in neutrino mass over momentum transfer, the
decay amplitude is proportional to the effective neutrino mass
\begin{equation}
m_{\beta\beta}=|\sum_jm_j(V^{\rm C}_{1j})^2|
=|V_{1n}|^2|m_-(c_m)^2+m_+(\pm is_m)^2|
=0
\end{equation}
Although the Majorana nature of neutrinos in principle allows the decay to occur,
it is highly suppressed in the considered model.

Finally, we want to study whether the model can accommodate the neutrino
oscillation data excluding LSND $\cite{note}$.
This would require at least two mass squared
differences and three mixing angles. From our above analysis,
we know that with a single sterile neutrino there are only two free mixing
angles for three generations. Although this is augmented by an effective mixing
angle formed by the two neutrino masses, the chance to accommodate all data
looks small. Our analysis below makes this claim stronger. As a first attempt,
it is enough to consider the data that may be reasonably well described by
vacuum neutrino oscillations, that is, those by KamLAND, K2K and CHOOZ.
These experiments produce and detect neutrinos by CC interactions.
The amplitude for the whole process that produces a neutrino and a charged
lepton $\bar{\ell}_{\alpha}$ at the source and detects a charged lepton
$\ell_{\beta}$ at the detector is proportional to
$\sum_jV^{{\rm C}*}_{\alpha j}V^{\rm C}_{\beta j}\exp[-im_j^2L/(2E)]$, where
$L$ is the source-detector distance and $E$ the energy of a relativistic
neutrino. The probability is
\begin{equation}
P(\alpha\to\beta)=\displaystyle\delta_{\alpha\beta}
-4\sum_{i>j}{\rm Re}(Q_{\alpha\beta;ij})\sin^2\frac{\Delta_{ij}L}{4E}
+2\sum_{i>j}{\rm Im}(Q_{\alpha\beta;ij})\sin\frac{\Delta_{ij}L}{2E}
\end{equation}
where
$\Delta_{ij}=m_i^2-m_j^2$ and
$Q_{\alpha\beta;ij}=V^{{\rm C}*}_{\alpha i}V^{\rm C}_{\beta i}
V^{{\rm C}*}_{\beta j}V^{\rm C}_{\alpha j}$.

The above formula simplifies considerably in our model due to the special
structure of $V^{\rm C}$ and the existence of only three different masses.
Eqs. $(\ref{eq_factor},\ref{eq_u})$ give
\begin{equation}
V^{\rm C}_{\alpha j}V^{{\rm C}*}_{\beta j}
=\sum_{\gamma=1}^{n-1}V_{\alpha\gamma}V_{\beta\gamma}^*\delta_{j\gamma}
+V_{\alpha n}V_{\beta n}^*(c_m^2\delta_{jn}+s_m^2\delta_{j,n+1})
\end{equation}
where the $i$ factor in $U$ drops out already. Taking into account the
three different masses, we only need the following quantities
\begin{equation}
\begin{array}{rcl}
\displaystyle\sum_{j=1}^{n-1}Q_{\alpha\beta;n,j}
&=&c_m^2|V_{\alpha n}|^2(\delta_{\alpha\beta}-|V_{\beta n}|^2),\\
\displaystyle\sum_{j=1}^{n-1}Q_{\alpha\beta;n+1,j}
&=&s_m^2|V_{\alpha n}|^2(\delta_{\alpha\beta}-|V_{\beta n}|^2),\\
Q_{\alpha\beta;n+1,n}&=&c_m^2s_m^2|V_{\alpha n}V_{\beta n}|^2,
\end{array}
\end{equation}
where unitarity of $V$ is used in the first two equalities. They are real
even without using the reality of $V$ in eq. $(\ref{eq_v})$. Thus the
probability for the associated process with $\bar{\ell}_{\alpha},~\ell_{\beta}$
replaced by $\ell_{\alpha},~\bar{\ell}_{\beta}$ is the same. Denoting
$x_{\pm}=m_{\pm}^2L/(4E)$, we obtain
\begin{equation}
\begin{array}{rcl}
P(\alpha\to\beta)&=&\displaystyle\delta_{\alpha\beta}
-4|V_{\alpha n}|^2(\delta_{\alpha\beta}-|V_{\beta n}|^2)
(c_m^2\sin^2x_-+s_m^2\sin^2x_+)\\
&&\displaystyle
-4|V_{\alpha n}V_{\beta n}|^2c_m^2s_m^2\sin^2(x_+-x_-)
\end{array}
\end{equation}
Note that the probability does not normalize to unity $\cite{sv80a}$:
\begin{equation}
\displaystyle\sum_{\beta}P(\alpha\to\beta)=
1-4|V_{\alpha n}|^2(c_ms_m)^2\sin^2(x_+-x_-)
\end{equation}
This occurs because the mixing matrix appearing in the amplitude is not the
one relating weak and mass eigenstates of neutrinos which is always unitary
and would guarantee unity normalization,
but the one appearing in CC interactions which is not unitary
in the current model so that the sterile degree of freedom gets effectively
lost in the sum.

The KamLAND and CHOOZ are reactor $\bar{\nu}_e$ disappearance experiments at
$L/E\sim (20-50)\times 10^3 {\rm ~km/GeV}$ and
$L/E\sim 333 {\rm ~km/GeV}$ respectively, and are thus potentially sensitive
to small and large mass squared differences. The K2K experiment observes
accelerator $\nu_{\mu}$ disappearance at $L/E\sim 200 {\rm ~km/GeV}$,
close to the range at CHOOZ though in a different channel.
That KamLAND and K2K observed deficits and spectral distortions implies that
there would be two well-separated mass squared differences.
Then, the null result by CHOOZ would be interpreted by some small mixing
parameter. There are two ways to arrange this, either
(1) $m_+^2\sim m_-^2\gg m_+^2-m_-^2$
or
(2) $m_+^2\gg m_-^2$, implying correspondingly
(1) $c_m\sim s_m\sim 1/\sqrt{2}$
or
(2) $c_m\sim 1,~s_m\sim 0$.
In case (2), oscillations in the larger mass squared difference, i.e., in
$x_+\sim x_+-x_-$, are highly suppressed in all channels, thus incapable of
accommodating the K2K data. In case (1), we have
\begin{equation}
P(\alpha\to\alpha)\sim\displaystyle
1-4|V_{\alpha n}|^2(1-|V_{\alpha n}|^2)\sin^2x_+
-|V_{\alpha n}|^4\sin^2(x_+-x_-)
\end{equation}
For K2K and CHOOZ, the last term can be ignored so that their results imply
that
$4|V_{2n}|^2(1-|V_{2n}|^2)\sim 1$ and
$4|V_{1n}|^2(1-|V_{1n}|^2)\sim 0$.
Since unitarity of $V$ demands that $|V_{1n}|^2+|V_{2n}|^2\le 1$,
the combined solution is,
$|V_{2n}|^2\sim 1/2$ and $|V_{1n}|^2\sim 0$.
But the latter is rejected by the KamLAND data.

\section{Conclusion}

We have investigated the leptonic mixing matrices in SM augmented by one sterile
neutrino. We found that the mixing matrix $V^{\rm C}$ in CC interactions takes a factorized
form, with one factor $U$ describing the mixing in the subsystem of the two massive
neutrinos and the other $V$ the mixing of the massless neutrinos with the subsystem.
We showed that this arises from the texture zero in the neutrino mass matrix
that is protected by gauge symmetry. The matrix $U$ is completely fixed by the two
masses, while the factorization makes it possible to remove all phases in $V$,
thus leaving us with $(n-1)$ free physical mixing angles in $V^{\rm C}$ for $n$
generations. The factorization also determines uniquely the mixing matrix $V^{\rm N}$
in NC neutrino interactions in terms of the masses so that off-diagonal interactions
occur only between the massive neutrinos. Both CC and NC interactions automatically
conserve CP. We also considered some phenomenological results from the exposed
structure in $V^{\rm C}$. The effective neutrino mass in tritium $\beta$ decay is
essentially sensitive to the Dirac mass in the model, while the leading contribution
to neutrinoless double $\beta$ decay vanishes. The difficulty with this simple model
is that, even with any number of generations, it cannot accommodate the vacuum neutrino
oscillation data coming from KamLAND, K2K and CHOOZ. A way out might be to introduce
two sterile neutrinos, which would bring in more free parameters into the model.
While the involvement of sterile neutrinos is not favored in solar and atmospheric
data, it might still be phenomenologically viable by relaxing the tension between
the LSND and other short baseline experiments, if either of the two can be
unambiguously confirmed by MiniBooNE.



\end{document}